\documentclass{llncs}

\usepackage[ruled,vlined]{algorithm2e}
\usepackage{stmaryrd}

\usepackage{calc}
\usepackage{etex}
\usepackage{times}

\newenvironment{itemize*}%
  {\begin{itemize}%
    \setlength{\itemsep}{3pt}%
    \setlength{\parskip}{0pt}}%
  {\end{itemize}}

\usepackage{amsfonts}
\usepackage{amssymb}
\usepackage{textcomp}
\usepackage{colortbl}
\usepackage{xspace}
\usepackage{alltt}
\usepackage{color}
\usepackage{subfig}
\usepackage{listings}

\usepackage{url}
\usepackage[colorlinks=true,
            urlcolor=blue,
            citecolor=blue,
            linkcolor=blue
            ]{hyperref}

\usepackage{tikz}
\usetikzlibrary{arrows,shapes,shadows}

\usepackage{verbatim} 

\usepackage{mathtools}

\usepackage{graphicx}

\usepackage[T1]{fontenc}
\usepackage{enumitem}
\usepackage{fancybox}

\newcommand{\mypara}[1]{\vspace{5pt}\noindent\textbf{#1.}}
\newcommand{\myparax}[1]{\vspace{5pt}\noindent\textbf{#1}}

\newcommand{\Figref}[1]{Figure\,\ref{#1}}
\newcommand{\figref}[1]{Fig.\,\ref{#1}}

\newcommand{\la}{\ensuremath{\leftarrow}}

\newcommand{\solved}[1]{\ensuremath{#1^\lambda}}

\newcommand{\xgreen} {{\ensuremath{\textcolor{DarkGreen}{\mathsf{green}}}}}
\newcommand{\xred}     {{\ensuremath{\textcolor{DarkRed}{\mathsf{red}}}}}
\newcommand{\xyellow}{{\ensuremath{\textcolor{DarkYellow}{\mathsf{yellow}}}}}

\newcommand{\green} {{\ensuremath{\textcolor{DarkGreen}{\mathsf{g}}}}}
\newcommand{\red}     {{\ensuremath{\textcolor{DarkRed}{\mathsf{r}}}}}
\newcommand{\yellow}{{\ensuremath{\textcolor{DarkYellow}{\mathsf{y}}}}}

\newcommand{\won} {{\ensuremath{\textcolor{DarkGreen}{\mathsf{W}}}}\xspace}
\newcommand{\lost}     {{\ensuremath{\textcolor{DarkRed}{\mathsf{L}}}}\xspace}
\newcommand{\drawn}{{\ensuremath{\textcolor{DarkYellow}{\mathsf{D}}}}\xspace}

\definecolor{DarkGreen}{rgb}{0,0.45,0}
\definecolor{DarkRed}{rgb}{0.8,0,0}
\definecolor{DarkYellow}{rgb}{0.6,0.6,0}
\definecolor{DarkGray}{rgb}{0.2,0.2,0.2}

\newcommand{\wonPos}[1]{\textcolor{DarkGreen}{{\emph{\textbf{#1}}}}}
\newcommand{\lostPos}[1]{\textcolor{DarkRed}{{\textbf{\emph{#1}}}}}
\newcommand{\drawnPos}[1]{\textcolor{DarkYellow}{{\emph{\textbf{#1}}}}}

\begin{document}

\title{Games and Argumentation: Time for a Family Reunion!}

\author{Bertram Lud{\"a}scher \and Yilin Xia}

\institute{
School of Information Sciences\\
  University of Illinois, Urbana-Champaign\\
  \email{\{ludaesch,yilinx2\}@illinois.edu}
}

\maketitle

\newcommand{\pos}[1]{\ensuremath{\mathsf{#1}}}

\newcommand{\winmove}{\pos{win}-\pos{move}}
\newcommand{\AF}{\ensuremath{\mathsf{AF}}}
\newcommand{\WM}{\ensuremath{\mathsf{WM}}}

\section{Introduction}


Consider the following two single-rule programs $P_\mathsf{AF}$ and $P_\mathsf{G}$ that deal with the status of
abstract arguments and with positions in a game graph, respectively:

\begin{equation}
  \pos{defeated}(X) \la \pos{attacks}(Y, X), \neg \, \pos{defeated}(Y). \tag{$P_\mathsf{AF}$}
\end{equation}

\begin{equation}
  \pos{win}(X) \la \pos{move}(X, Y), \neg \, \pos{win}(Y). \tag{$P_\mathsf{G}$}
\end{equation}

\newcommand{\ren}{\ensuremath{\rightleftharpoons}}

$P_\mathsf{AF}$ states that an argument $X$ is \emph{defeated} in an {argumentation framework} AF if
it is \emph{attacked} by an argument $Y$ that is accepted, i.e., {not} defeated. Conversely, $P_\mathsf{G}$ states that in a
{game}  a position $X$ is \emph{won} if there is a \emph{move} to a position $Y$ that is {not}
won (by the opponent).  Both logic rules can be seen as close relatives, even ``identical
twins'',\footnote{The rules $P_\mathsf{AF}$ and $P_\mathsf{G}$ are syntactic variants of each other:   swap relations $\pos{defeated}\ren\pos{win}$ and
  $\pos{attacks}\ren\pos{move}^{-1}$ (the direction of edges is reversed in \pos{move}$^{-1}$).} and each rule has
received considerable attention in the past by different communities:


The first rule $(P_\mathsf{AF})$ constitutes an ``argument processing unit'' (APU) that---together with a suitable semantics---lies at the heart of
Dung's abstract argumentation theory~\cite{dung1995acceptability}, a seminal work that spawned a large body of research,
including families of models, semantics, tools, and applications of abstract (and structured) argumentation
\cite{baroni_handbook_2018,baroni_acceptability_2020,besnard_logical_2020}.

The second rule ($P_\mathsf{G}$) has played a key role in the logic programming and non\-monotonic reasoning community in their quest to find the ``right'' semantics for rules with recursion through
negation (i.e., which are \emph{not stratified}\footnote{A \emph {stratified} logic program $P$ \cite {apt_towards_1988} can use both (positive) recursion and negation, but only in a ``layered''  manner, i.e., the rule-goal graph of $P$ must not contain negative cycles.}), and in database theory.

For non-stratified logic programs such as $P_\mathsf{AF}$ and $P_\mathsf{G}$, two declarative semantics
emerged as the most popular, i.e., the \emph{stable model semantics}
\cite{gelfond_stable_1988} 
and the \emph{well-founded semantics} \cite{van1991well}. For the latter, the \pos{win}-\pos{move}
game $P_\mathsf{G}$ has been the poster-child example because its unique three-valued model assigns
\emph{true}, \emph{false}, and \emph{undefined} to $\pos{win}(x)$ iff a position $x$ in the given
game graph is \emph{won}, 
\emph{lost}, or \emph{drawn}, respectively. In other words, $P_\mathsf{G}$ \emph{solves games} and thus can
be viewed as a ``game processing unit'' (GPU) when used with an engine that computes the
well-founded semantics.  
Similarly, the rule $P_\mathsf{AF}$ is an APU  that can solve
 argumentation frameworks: The well-founded model of $P_\mathsf{AF}$ yields \emph{grounded
  extensions} and \emph{grounded labelings} \cite{caminada_logical_2009,modgil2009proof}, where an
argument $x$ is \emph{defeated} (label = \texttt{out}), \emph{accepted} (label = \texttt{in}), or
\emph{undecided} (label = \texttt{undec}) iff $\pos{defeated}(x)$ is \emph{true}, \emph{false}, and
\emph{undefined} in the well-founded model, respectively.


Although close connections between formal argumentation on one hand, and logic programming, nonmonotonic reasoning, and games on the
other have been known for a long time \cite{dung1995acceptability,modgil2009proof,caminada2015equivalence,baroni_handbook_2018}, the overlap and cross-fertilization between these and some other
communities (e.g., database theory) appears to be smaller than one might expect. There seems to be no work, e.g., that discusses
$P_\mathsf{AF}$ and $P_\mathsf{G}$ in the same paper, despite (or because of?) the fact that these rules can be viewed as syntactic variants of the
same underlying query. In database theory, the \winmove\ query expressed by $P_\mathsf{G}$ has been used to study the \emph{expressive power}
of query languages \cite{kolaitis_expressive_1991,FKL97} and to develop a unified \emph{provenance model} that can explain the
presence and absence of query answers \cite{kohler_first-order_2013,lee_sql_middleware_2017}.

The game-theoretic notions and concepts developed in these and other database and game-theory papers
\cite{fraenkel1997combinatorial,flum2000games} seem to carry over to argumentation theory and may 
lead to new insights and results there. Conversely, related notions studied in argumentation theory may carry over to  database theory and applications thereof.
The purpose of this short paper is therefore to foster a ``family reunion'' of sorts 
with the goal of developing new insights and findings through cross-fertilization, i.e., by transferring concepts, ideas, and results between communities.




\section{Some Sightings of the \pos{Win}-\pos{Move} Query in Database Theory} 
We recall first some (possibly lesser known) results about $P_\mathsf{G}$, the ``lost twin'' of $P_\mathsf{AF}$.

\mypara{Games vs Stratified Rules} During the late 1980s and through the 90s, the LP/NMR community developed and studied a number of
proposals for a canonical semantics for rules with recursion and negation. Veterans from that era will fondly recall examples of the
form $\{\pos{p} \la \neg \pos{q};~ \pos{q} \la \neg \pos{p}\}$, or the rather self-defeating (pun intended) 
$\{ \pos{p} \la \neg \pos{p}\}$.  Proponents of the \emph{stratified semantics} \cite{apt_towards_1988} simply ruled out such
unstratifiable programs, i.e., which exhibit recursion {through} negation. An earlier claim by \cite{chandra_horn_1985} suggested that
stratified rules express all of \textsc{Fixpoint} \cite{abiteboul1995foundations}, i.e., a large class of database queries with {PTIME}
data complexity.  As shown in \cite{kolaitis_expressive_1991}, however, the \textsc{Fixpoint} query that computes the positions for
which a player has a \emph{winning strategy}\footnote{Player\,I can force a win, no matter how Player\,II moves.} is \emph{not} expressible
by stratified rules, so stratified Datalog is strictly less expressive than \textsc{Fixpoint}.  On the other hand, well-founded
Datalog expresses all \textsc{Fixpoint} queries, and $P_\mathsf{G}$ computes the won, lost, and drawn positions in PTIME for any game given by
a finite \pos{move} graph.

\begin{figure}[th]
  \centering
\subfloat[What are the \emph{good moves}, say from position \texttt{e}?
Is \texttt{e} \emph{won} (or \emph{lost} or \emph{drawn}) and how can we \textbf{explain} it?]{
 \includegraphics[width=.28\columnwidth]{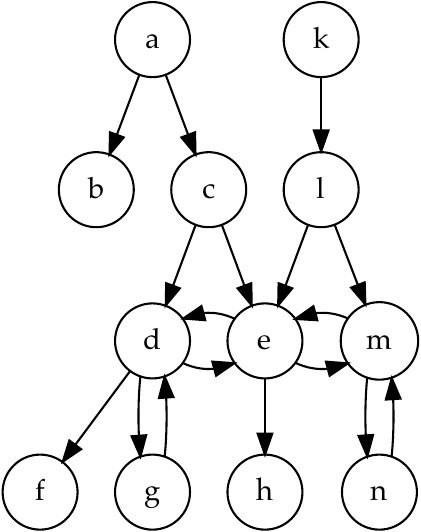}\label{fig-unsolved}}
\hfill
\subfloat[The \textbf{solved game} then reveals the answer: move
$\mathtt{e}{\to} \mathtt{h}$ is winning; the moves
$\mathtt{e}{\to}\mathtt{d}$ and $\mathtt{e}{\to}\mathtt{m}$ are not.]{
  \includegraphics[width=.28\columnwidth]{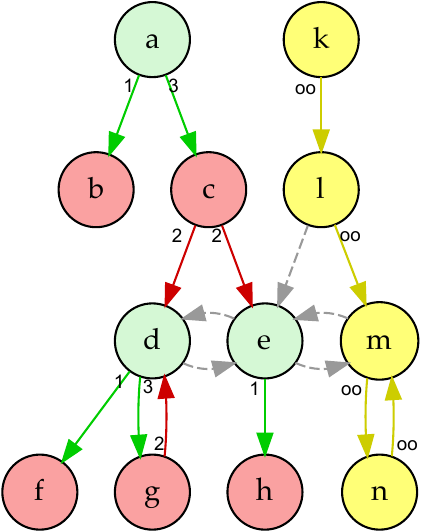}\label{fig-solved}}
\hfill
\subfloat[The colored (= \textbf{labeled}) \emph{argumentation framework} corresponds to the solved
game (b), \emph{mutatis mutandis}.]{
  \includegraphics[width=.28\columnwidth]{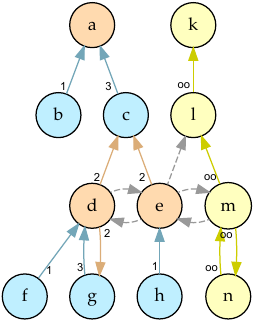}\label{fig-AF}}
\caption{\small The \pos{move} graph on the left defines a game $G=(V,E)$ with positions $V$ and
  moves $E$.\\
The \emph{solved game} \solved G in (b) is color-labeled: 
  positions 
  are either \emph{won} ({green}), \emph{lost} ({red}), or \emph{drawn} ({yellow}). This separates ``good'' moves
  (solid, colored) from ``bad'' ones (dashed, gray). The length $\ell$ of an edge
 $x{\stackrel{\ell}{\to}}y$
  indicates how quickly one can force a win, or how long one can delay a loss, with that move.  By
  reversing the \pos{move} edges, one obtains an \pos{attack} graph: its
  \emph{grounded labeling} in (c) shows arguments that are \texttt{out} (orange), \texttt{in} (blue), and \texttt{undec} (yellow).}
  \label{fig-unsolved-solved}
\end{figure}


\mypara{All You Need Is Game} Consider the rule $P_\mathsf{G}$ over a given \pos{move} graph, e.g., on the
left in \figref{fig-unsolved-solved}(a). $P_\mathsf{G}$ captures the essence of a 2-person game $G= (V,E)$
with {positions} $V$ and {moves} $E$. Initially, a pebble is placed on a start position, and
Player\,I starts to move. The players then take turns, moving the pebble along the edges of the graph
until a player runs out of moves,\footnote {For example, in chess this means: \emph{checkmate}!  There is no (legal) move left to play.} in
which case the opponent has won, i.e., a position $x\in V$ is \emph{won} for a player if there
\emph{exists} a move to a position $y$ that is \emph{lost} (i.e., not won) for the opponent.  If $y$
is objectively lost, this means that \emph{all} outgoing moves from $y$ lead to a position $z$ that
will again leave the first player in a won position. This alternation of quantifiers
($\exists x_1 \forall y_1 \exists x_2 \forall y_2 \cdots)$ lies at the core of the expressive power
of $P_\mathsf{G}$.  

The rule $P_\mathsf{G}$ turns out to be a \textbf{universal query engine}: every $n$-ary
\textsc{Fixpoint} query 
can be expressed in \emph{game normal form}
$P_\mathsf{G}: \pos{win}(\bar X) \la \pos{move}(\bar X, \bar Y), \neg \pos{win}(\bar Y)$, i.e., where $\bar X$ and
$\bar Y$ are $n$-tuples of variables, $P_\mathsf{G}$ is the only recursive rule, and $\pos{move}(\bar X, \bar
Y)$ is definable
via a quantifier-free formula over the input database \cite{FKL97}.

\mypara{Solving Games}
In \Figref{fig-unsolved-solved}(a) and (b), positions \pos b, \pos f, and \pos h are immediately
lost (red nodes): No moves are possible from sink nodes. Next we can infer that positions that have an outgoing move to a lost
position (for the opponent) are definitely won (green). Based on our initial determination that \pos
b, \pos f, and \pos h are lost, it then follows that \pos a, \pos d, and \pos e are won.\footnote{Similarly (\emph{mutatis mutandis}), in the argumentation framework \Figref{fig-unsolved-solved}(c), arguments \pos
b, \pos f, and \pos h are \emph{accepted} (not defeated) and can be labeled \texttt{in} because they are not attacked at all.}
What is the status of the remaining positions?  The status of \pos c is now determined since
\emph{all} outgoing moves from \pos c definitely end in a node that is
won for the opponent (\pos d and \pos e are already green),
so \pos c is objectively lost.  Solving a game thus proceeds by iterating the following two coloring
(or \emph{labeling}) rules in stages:\footnote{This method 
corresponds to the \emph{alternating fixpoint} procedure
\cite{van1993alternating} and 
to Algorithm 6.1 for computing the \emph{grounded labeling} of an argumentation framework in \cite{modgil2009proof}.}
\begin{itemize}
\item    Position $x$ is \emph{won} (green) if $\exists$ {move} $x\to y$ and position $y$  is lost (red) 
\item    Position $x$ is \emph{lost} (red) if $\forall$ {moves} $x\to y$, position $y$  is
  won (green)  
\end{itemize}
With each position $x$ we can associate its \emph{length}
\cite{kohler_first-order_2013}, i.e., the stage number when its color
first became known. Similarly, we can associate a length with each
move, indicating at what stage its \emph{type} (edge color) became
known: in \figref{fig-unsolved-solved}(b) edges into (\pos{red}) sinks
are winning moves (colored \pos{green}) and labeled with length = 1,
so \pos a, \pos d, \pos e and those edges to sink nodes all have
length = 1. In the next stage, all successors of \pos c are won, so
\pos c itself must be lost, and its length is 1 + the \emph{maximal} length
of any of its succcessors. Similarly, for won $x$, length($x$) = 1 +
the \emph{minimal} length of any lost successor, etc.

After a fixpoint is reached, all remaining uncolored nodes
correspond to \emph{drawn} positions and are colored \pos{yellow}. We set length = $\infty$ for
drawn positions, since neither player can force a win, but both can avoid losing by repeating moves
indefinitely.

\myparax{Solved Games Explain It All!} Solved games, e.g., \solved G in
\figref{fig-unsolved-solved}(b), have an intriguing property: node labels (colors) induce
different \emph{edge types}, which in turn can be used to \emph{explain why} a position is won,
lost, or drawn, respectively. \figref{fig-prov-edges} shows how edge types are determined from the
color-labels of incident vertices. These types, in turn, induce a downstream \emph{provenance subgraph} $G^\lambda_x$ which provides the \emph{justification} or \emph{explanation} for the status of any  $x\in V$. The provenance $G^\lambda_x$ of $x$ in the solved game $G^\lambda$ is the subgraph reachable from $x$ via certain \emph{regular path queries} (RPQs): The provenance of a {won} position
$x$ matches the RPQ $x.\xgreen.(\xred.\xgreen)^*$, {lost} positions match
$x.(\xred.\xgreen)^*$, and drawn positions match $x.\xyellow^+$.
Similarly, an argument's status in the grounded argumentation framework in \figref{fig-unsolved-solved}(c) can be explained by an RPQ-definable subgraph, \emph{mutatis mutandis}. Unless this is known by a different name in argumentation frameworks, this seems
to be a new result for grounded extensions.

The fact that (i) solved games expose their own provenance (= result
\emph{explanations}), and (ii) all \textsc{First-Order} queries $Q$ have a
natural encoding as query evaluation games $G_Q$, were
combined in \cite{kohler_first-order_2013} to develop
\emph{first-order provenance games}, a unified framework for
\emph{why}-, \emph{how}-, and \emph{why-not} provenance.
The framework can explain {how} a query result was derived, and why
some  results are {missing} from an answer $A=Q(D)$ (see also
\cite{lee_sql_middleware_2017}). Roughly speaking, two players argue
(via a query evaluation game) whether or not tuple $t\in A$, for a given database $D$. The provenance of $t$ corresponds to a subgraph in the solved game $G_{Q(D)}^\lambda(t)$ and to the winning strategies for the claim $t\in A$.




\newcommand{\na}{\textcolor{gray}{\emph{n/a}}}
\begin{figure}[t]
  \centering
  \renewcommand{\arraystretch}{1.5}
~\hfill
  {\small\begin{tabular}[b]{r|c|c|c}
      & \wonPos{$y$ won}  (\won) &  \drawnPos{$y$ drawn} (\drawn)&
       \lostPos{$y$ lost} (\lost)\\
      \hline
      \wonPos{$x$ won} (\won) & \textcolor{gray}{\emph{{bad}}} & \textcolor{gray}{\emph{{bad}}} & 
      \textcolor{DarkGreen}{\textbf{\green: \emph{winning}}} \\  
      \hline
       \drawnPos{$x$ drawn} (\drawn) & \textcolor{gray}{\emph{{bad}}} & \textcolor{DarkYellow}{\textbf{\yellow: \emph{drawing}}}  &\na\\  
      \hline
       \lostPos{$x$ lost} (\lost) &
      \textcolor{DarkRed}{\textbf{\red: \emph{delaying}}}  &\na & \na \\  
    \end{tabular}}
\hfill\hfill\hfill\hfill
  \includegraphics[width=.38\columnwidth]{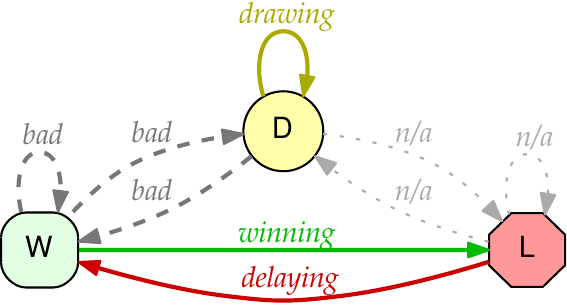}
\hfill~
  \caption{\small Depending on node labels, moves $x\to y$ are 
    either \emph{winning} (or \pos{green}) ($\won \stackrel{\green}{\leadsto} \lost$),
    \emph{delaying} (or \pos{red}) ($\lost \stackrel{\red}{\leadsto} \won$), or
    \emph{drawing} (or \pos{yellow}) ($\drawn\stackrel{\yellow}{\leadsto}\drawn$). All
    other moves are either ``\emph{bad}'' (allowing the opponent to
    improve the outcome), or cannot exist (\na) due to the nature of the game: e.g., if $x$ is lost,
    then there are only delaying moves (i.e., ending in won positions
    $y$ for the opponent).}
\label{fig-prov-edges}
\end{figure}
\renewcommand{\arraystretch}{1}

\newcommand{\wfg}{\ensuremath{\mathcal{M}^{w}_\mathsf{G}}}
\newcommand{\wfaf}{\ensuremath{\mathcal{M}^{w}_\mathsf{AF}}}

\section{The Next Move: Strengthening Family Ties}

This brief exposition of a few results from database theory for 
the \winmove\ query $P_\mathsf{G}$ should look familiar to researchers in formal argumentation. Clearly, there are many direct correspondences, but
there are also slightly different notions that seem to warrant
further inspection and investigation, and that could lead to new insights and results.

Our starting point was the straightforward link between $P_\mathsf{AF}$ and $P_\mathsf{G}$: Twin rules that have their distinct histories and applications in separate communities, but that haven't been studied together, at least to the best of our knowledge. 
Under the well-founded semantics, the solved game $G^\lambda$ (with
its additional structure and ``built-in'' provenance) corresponds to
the grounded labeling of an argumentation framework~\cite{caminada2006issue,modgil2009proof}. The  additional provenance structure  induced by edge types   (not all edges are ``created equal'') suggests a corresponding new structure for
AF.

Here are a few more propositions and conjectures, aimed at fostering new collaborations between our different communities:

\begin{itemize}
\item The well-founded model $\wfg$ of $P_\mathsf{G}$ is isomorphic to the well-founded model $\wfaf$ of $P_\mathsf{AF}$ via a natural mapping (cf.~\Figref{fig-unsolved-solved}(b) and (c)). What about other logical semantics, e.g., stable models?  The resulting answer sets are  useful in the analysis of AFs, but do they have natural and intuitive interpretations for games?  After all, it is the well-founded semantics that yields the canonical model for \winmove\ games.
\item Color labels of solved games correspond to grounded labelings of AF
  (\Figref{fig-unsolved-solved}).  The \emph{length} of game positions in \wfg\
  \cite{kohler_first-order_2013} (see also \cite{FKL97}) corresponds to the \emph{min-max numbering} of arguments in \wfaf\
  \cite{baroni18:_abstr_argum_framew_their_seman}, and 
   also appears as a byproduct in the alternating fixpoint computation \cite{van1993alternating} when solving a game via the $P_\mathsf{G}$ rule.\footnote{A similar stage/state number is used 
  for analyzing the provenance of Datalog queries  \cite{kohler2012declarative}.} 
  \item The \emph{characteristic function} \cite{dung1995acceptability} in argumentation frameworks is  closely related to  \emph{strategy functions} \cite{flum2000games}, \emph{winning strategies} \cite{FKL97} and the \emph{unattacked} operator $U_{\Theta}$ \cite{bogaerts2015grounded}. What are the precise correspondences? What results might be transferable? 
  \item There is a plethora of variants of argument/dialogue games in argumentation theory, in addition to the basic \pos{win}-\pos{move} games used in databases  \cite{kolaitis_expressive_1991,FKL97,FKL00-TCS} and game theory \cite{fraenkel1997combinatorial,flum2000games}. Has someone classified this ``zoo'' of game variants before?
 \item The \emph{decomposition theorems} for graph kernels \cite{fraenkel1997combinatorial,flum2000games}
   directly apply to games and thus carry over to argumentation frameworks. Has this been studied before?
 \end{itemize}

\noindent We invite feedback and welcome collaboration opportunities on these and similar questions. An open source demonstration using Jupyter notebooks, including the example from \Figref{fig-unsolved-solved}, is available  \cite{xia_games-and-argumentation_2023}. We plan to evolve and expand these notebooks as teaching materials for some of our undergraduate and graduate courses, covering knowledge representation \& reasoning, information modeling, and database theory.

\mypara{Acknowledgments} The authors thank Shawn Bowers for his detailed comments and suggestions on an earlier draft. Work supported in part by NSF/OAC-2209628 (TRACE).




\bibliographystyle{splncs03}
\bibliography{argumentation_games}

\begin{thebibliography}{10}
\providecommand{\url}[1]{\texttt{#1}}
\providecommand{\urlprefix}{URL }

\bibitem{abiteboul1995foundations}
Abiteboul, S., Hull, R., Vianu, V.:
  \href{http://webdam.inria.fr/Alice/}{Foundations of Databases}.
  Addison-Wesley (1995)

\bibitem{apt_towards_1988}
Apt, K.R., Blair, H.A., Walker, A.:
  \href{https://www.sciencedirect.com/science/article/pii/B9780934613408500063}{Towards
  a {Theory} of {Declarative} {Knowledge}}. In: Minker, J. (ed.) Foundations of
  {Deductive} {Databases} and {Logic} {Programming}, pp. 89--148 (1988)

\bibitem{baroni18:_abstr_argum_framew_their_seman}
Baroni, P., Caminada, M., Giacomin, M.: Abstract argumentation frameworks and
  their semantics. In: Handbook of {Formal} {Argumentation}
  \cite{baroni_handbook_2018}, chap.~4, pp. 159--236

\bibitem{baroni_handbook_2018}
Baroni, P., Gabbay, D., Giacomin, M., Torre, L.v.d.:
  \href{https://philpapers.org/rec/BARHOF}{Handbook of {Formal}
  {Argumentation}}. London, England: College Publications (2018)

\bibitem{baroni_acceptability_2020}
Baroni, P., Toni, F., Verheij, B.:
  \href{https://content.iospress.com/articles/argument-and-computation/aac200901}{On
  the acceptability of arguments and its fundamental role in nonmonotonic
  reasoning, logic programming and n-person games: 25 years later}. Argument \&
  Computation  11(1-2),  1--14 (Jan 2020)

\bibitem{besnard_logical_2020}
Besnard, P., Cayrol, C., Lagasquie-Schiex, M.C.:
  \href{https://content.iospress.com/articles/argument-and-computation/aac190476}{Logical
  theories and abstract argumentation: {A} survey of existing works}. Argument
  \& Computation  11(1-2),  41--102 (Jan 2020)

\bibitem{bogaerts2015grounded}
Bogaerts, B., Vennekens, J., Denecker, M.:
  \href{https://doi.org/10.1016/j.artint.2015.03.006}{Grounded Fixpoints and
  Their Applications in Knowledge Representation}. Artificial Intelligence
  224,  51--71 (Jul 2015)

\bibitem{caminada2006issue}
Caminada, M.: \href{https://doi.org/10.1007/11853886_11}{On the {{Issue}} of
  {{Reinstatement}} in {{Argumentation}}}. In: Logics in {{Artificial
  Intelligence}}. pp. 111--123. LNAI 4160, {Springer} (2006)

\bibitem{caminada2015equivalence}
Caminada, M., S{\'a}, S., Alc{\^a}ntara, J., Dvo{\v r}{\'a}k, W.:
  \href{https://doi.org/10.1016/j.ijar.2014.12.004}{On the Equivalence between
  Logic Programming Semantics and Argumentation Semantics}. International
  Journal of Approximate Reasoning  58,  87--111 (Mar 2015)

\bibitem{caminada_logical_2009}
Caminada, M.W.A., Gabbay, D.M.:
  \href{https://doi.org/10.1007/s11225-009-9218-x}{A {Logical} {Account} of
  {Formal} {Argumentation}}. Studia Logica  93(2),  109 (Nov 2009)

\bibitem{chandra_horn_1985}
Chandra, A.K., Harel, D.:
  \href{http://www.sciencedirect.com/science/article/pii/0743106685900020}{Horn
  clause queries and generalizations}. The Journal of Logic Programming  2(1),
  1--15 (1985)

\bibitem{dung1995acceptability}
Dung, P.M.:
  \href{https://www.sciencedirect.com/science/article/pii/000437029400041X}{On
  the Acceptability of Arguments and Its Fundamental Role in Nonmonotonic
  Reasoning, Logic Programming and n-Person Games}. Artificial Intelligence
  77(2),  321--357 (Sep 1995)

\bibitem{flum2000games}
Flum, J.: \href{https://doi.org/10.1023/A:1006463732291}{Games, {{Kernels}},
  and {{Antitone Operations}}}. Order  17(1),  61--73 (Mar 2000)

\bibitem{FKL97}
Flum, J., Kubierschky, M., Lud{\"a}scher, B.:
  \href{https://doi.org/10.1007/3-540-62222-5_40}{Total and Partial
  Well-Founded Datalog Coincide}. In: Afrati, F., Kolaitis, P. (eds.) ICDT,
  Delphi. pp. 113--124. LNCS 1186, {Springer} (1997)

\bibitem{FKL00-TCS}
Flum, J., Kubierschky, M., Lud{\"a}scher, B.:
  \href{https://www.sciencedirect.com/science/article/pii/S0304397599002224}{Games
  and Total {{Datalog}}{$\lnot$} Queries}. Theoretical Computer Science
  239(2),  257--276 (May 2000)

\bibitem{fraenkel1997combinatorial}
Fraenkel, A.: \href{https://doi.org/10.37236/1325}{Combinatorial Game Theory
  Foundations Applied to Digraph Kernels}. Electronic Journal of Combinatorics
  4(2 R),  1--17 (1997)

\bibitem{gelfond_stable_1988}
Gelfond, M., Lifschitz, V.:
  \href{http://www.cs.utexas.edu/users/ai-lab?gel88}{The Stable Model Semantics
  for Logic Programming}. In: ILPS. pp. 1070--1080 (1988)

\bibitem{kohler2012declarative}
K{\"o}hler, S., Lud{\"a}scher, B., Smaragdakis, Y.:
  \href{https://doi.org/10.1007/978-3-642-32925-8_12}{Declarative Datalog
  Debugging for Mere Mortals}. In: Datalog 2.0: Datalog in Academia and
  Industry. pp. 111--122. LNCS 7494 (2012)

\bibitem{kohler_first-order_2013}
K\"ohler, S., Lud\"ascher, B., Zinn, D.:
  \href{https://doi.org/10.1007/978-3-642-41660-6_20}{First-order provenance
  games}. In: Tannen, V., Wong, L., Libkin, L., Fan, W., Tan, W.C., Fourman,
  M.P. (eds.) In Search of Elegance in the Theory and Practice of Computation.
  LNCS, vol. 8000, pp. 382--399. Springer (2013)

\bibitem{kolaitis_expressive_1991}
Kolaitis, P.G.:
  \href{http://www.sciencedirect.com/science/article/pii/089054019190059B}{The
  expressive power of stratified logic programs}. Information and Computation
  90(1),  50--66 (1991)

\bibitem{lee_sql_middleware_2017}
Lee, S., K\"ohler, S., Lud\"ascher, B., Glavic, B.:
  \href{https://doi.org/10.1109/ICDE.2017.105}{A {SQL}-Middleware Unifying Why
  and Why-Not Provenance for First-Order Queries}. In: ICDE. pp. 485--496
  (2017)

\bibitem{modgil2009proof}
Modgil, S., Caminada, M.:
  \href{https://doi.org/10.1007/978-0-387-98197-0_6}{Proof {{Theories}} and
  {{Algorithms}} for {{Abstract Argumentation Frameworks}}}. In: Simari, G.,
  Rahwan, I. (eds.) Argumentation in {{Artificial Intelligence}}, pp. 105--129
  (2009)

\bibitem{van1993alternating}
Van~Gelder, A.: \href{https://doi.org/10.1016/0022-0000(93)90024-Q}{The
  alternating fixpoint of logic programs with negation}. Journal of Computer
  and System Sciences  47(1),  185--221 (1993)

\bibitem{van1991well}
Van~Gelder, A., Ross, K.A., Schlipf, J.S.:
  \href{http://doi.acm.org/10.1145/116825.116838}{The Well-founded Semantics
  for General Logic Programs}. Journal of the ACM  38(3),  619--649 (1991)

\bibitem{xia_games-and-argumentation_2023}
Xia, Y., Lud\"ascher, B.: Games and argumentation demo repository (Aug 2023),
  \href{https://github.com/idaks/Games-and-Argumentation/tree/main/XLoKR-2023}{github.com/idaks/Games-and-Argumentation/tree/main/XLoKR-2023}

\end{thebibliography}

\end{document}